\documentstyle[12pt]{article}
\input epsf.sty
\newcommand{\be}{\begin{equation}}
\newcommand{\ee}{\end{equation}}
\newcommand{\bea}{\begin{eqnarray}}
\newcommand{\eea}{\end{eqnarray}}

\newcommand{\eps}{\epsilon}

\newcommand{\kp}{\kappa}
\newcommand{\lm}{\lambda}
\newcommand{\Lm}{\Lambda}

\newcommand{\nn}{\nonumber}

\begin{document}
\title{Exact Inflationary Models 
and Consequences}
\author{Apostolos Kuiroukidis\thanks{E-mail: 
kuirouki@astro.auth.gr}\\
and\\
D. B. Papadopoulos\thanks{E-mail: 
papadop@astro.auth.gr}
\\
Department of Physics, \\
\small Section of Astrophysics, Astronomy and Mechanics, \\
\small Aristotle University of Thessaloniki, \\
\small 54006 Thessaloniki, GREECE}

\maketitle
\begin{abstract}
New exact inflationary solutions are presented in the scalar field 
theory, minimally coupled to gravity, with a potential term. 
No use is made of the {\it slow rollover approximation}. 
The scale factors are completely nonsingular and the transition 
from the inflationary phase to the deccelerating phase is smooth. 
Moreover, in one of these models, asymptotically one has transition 
to the matter dominated Universe. 
\end{abstract}

\newpage 
\section*{I. Introduction}\
Inflation was historically introduced to solve the problem of monopoles 
that could be largely produced in the early Universe. It was subsequently 
elaborated into other forms such as the {\it chaotic inflation} scenario 
[1-3]. Inflation consists of a short period of accelerated superluminal 
expansion of the early Universe, at the end of which the transition to 
the standard big bang model should occur [4]. This apparently solves the 
problem of the flatness and smoothness of the Universe over such a large 
scale of distances [5-6]. 

In the scalar field driven inflationary scenario, the matter content 
of the Universe has the equation of state of the quantum vaccum, 
$P=-\rho $ (with $P, \rho $ the pressure and energy density of matter). 
The evolution is regarded as the "rolling" of the value 
of the field which is minimally coupled to gravity, in the presence 
of the scalar field potential. This potential is motivated usually 
from particle physics arguments. This is effectivelly a period of
supercooling of the Universe where in fact through a 
phase transition the field rolls to its true vacuum state [3,5]. 

The equations of motion for the scale factor of the Universe and
the field are usually treated in the so called {\it slow rollover} 
approximation. This is effectivelly equivalent to the flatness of the 
potential around the "false vacuum" value of the scalar field. 
However this is not always necessary. Motivated by this we present 
a class of solutions to these equations that are treated exactly. 
No use is made of the approximation and the various terms in the  
scalar field equation of motion are present throughout. 
These matters are treated in sections II, and III, while in section IV 
the numerical treatment of the problem is given. 
These models are nonsingular for all the positive values of the
cosmic time and in one of them occurs, in a natural manner, 
the asymptotic limit of a matter dominated Universe. 

\section*{II. Inflation in FLRW Models}\
In this section we review some basic facts about inflationary 
models. Then the first exact solution is presented which 
is generalized in the next section. \\
The metric is a member of the Friedman-Lemaitre-
Robertson-Walker class with flat spatial sections 
\be 
ds^{2}=-dt^{2}+a^{2}(t)(dx^{2}+dy^{2}+dz^{2}),
\ee 
where $a(t)$ is the scale factor of the Universe. Inflation is 
essentially scalar field dynamics for a scalar field $\phi $, 
minimally coupled to gravity. The energy density and pressure 
are given by 
\bea 
\rho &=&\frac{(\dot{\phi })^{2}}{2}+V\\
P&=&\frac{(\dot{\phi })^{2}}{2}-V,
\eea 
where $V(\phi )$ is the potential and $\phi (t)$ is assumed to 
depend solely on time. The Einstein's equations become 
\bea 
\frac{\ddot{a}}{a}&=&-\frac{4\pi G}{3}(\rho +3P)+
\frac{\Lm }{3}\\
H^{2}&\equiv &\left(\frac{\dot{a}}{a}\right)^{2}=
\frac{8\pi G}{3}\rho +\frac{\Lm }{3},
\eea 
with the possible presence of the cosmological constant, 
which will be neglected in what follows. The necessary 
condition for inflation is $\ddot{a}>0$. This corresponds 
to a negative pressure satisfying $P<-\rho /3$. 
Also we have the Klein-Gordon equation 
\be 
\label{pedio}
\ddot{\phi }+3H\dot{\phi }+\frac{dV(\phi )}{d\phi }=0.
\ee 
Only two of Eqs. (4)-(6) are truly independent. For example 
the first follows from the rest. Also the equation for the 
scalar field is equivalent to the energy conservation equation 
\be 
\dot{\rho }+3H(\rho +P)=0.
\ee 
We will consider only equations (5) and (6).\\
Different inflationary scenarios correspond to different 
choices of the scalar field potential $V(\phi )$ motivated 
by particle physics arguments. \\
The ansatz for the exact solution is 
\be 
\dot{\phi }=c_{0}V^{1/2}(\phi ),
\ee 
where for the present purposes $c_{0}$ is a constant. 
Then the solution to Eqs. (5)-(6) is given by 
\bea 
V(\phi )&=&V_{0}e^{\lm \phi }\nn \\
\lm &\equiv &-
\frac{c_{0}\sqrt{24\pi G}}{[1+c_{0}^{2}/2]^{1/2}}\; \; 
\; \; (c_{0}>0)\nn \\
\phi (t)&=&-\frac{2}{\lm }ln
\left[1-\frac{\lm }{2}c_{0}V_{0}^{1/2}t\right], \; \; 
\; \; (\phi (0)=0)\nn \\
H&=&\sqrt{\frac{8\pi G}{3}}
\left[1+\frac{c_{0}^{2}}{2}\right]^{1/2}V^{1/2}(\phi ).
\eea 
The condition for inflation $(\ddot{a}>0)$ is satisfied, 
from Eq. (4), for $c_{0}^{2}<1$. The scale factor becomes 
\be 
a(t)=a_{0}
\left[1-\frac{\lm }{2}c_{0}V_{0}^{1/2}t\right]^{E},
\ee 
where $a_{0}>0$ and $E$ is worked out to be 
$E=(1/3)+(2/3c_{0}^{2})$. This is completely 
nonsingular for all $t\geq 0$. The inflationary potential 
does not specify the equation of state for the scalar field, 
which has to be determined. In anticipation of the choice 
$0<c_{0}\ll 1$ that will be made, we obtain 
\be 
\frac{P}{\rho }=
\frac{(c_{0}^{2}/2)-1}
{(c_{0}^{2}/2)+1}\simeq -1,
\ee 
which is the equation of state of the quantum vacuum. 
The number of e-folds of inflation is given by 
\be 
N(\phi _{1}\rightarrow \phi _{2})\equiv 
\int _{t_{1}}^{t_{2}}Hdt=
\sqrt{\frac{8\pi G}{3}}\frac{1}{c_{0}}
\left[1+\frac{c_{0}^{2}}{2}\right]^{1/2}
(\phi _{2}-\phi _{1}).
\ee 
This can be made arbitrarily large provided that $c_{0}\ll 1$. 
We emphasize that there has not been imposed any condition on the 
{\it slow roll parameters} 
\bea 
\eps &=&\frac{1}{16\pi G}
\left(\frac{V^{'}}{V}\right)^{2}\nn \\
\eta &=&\frac{1}{8\pi G}
\frac{V^{''}}{V},
\eea 
i.e. $\eps ,|\eta |\ll 1$. Neither the "acceleration term" 
$\ddot{\phi }$, the "force term" $V^{'}\equiv dV/d\phi $, 
nor the "friction term" $3H\dot{\phi }$ have been neglected. 
Of course by choosing a suitably small constant $c_{0}$ one 
effectively mimics the conditions of slow rollover approximation 
but this is done in the context of an exact solution without 
any sort of approximation. 

\section*{III. Generalization of the Solution}\
The generalization of this solution comes from Eq. (8) when 
one considers the case $c_{0}=c_{0}(t)$. This function is 
proportional to the "canonical momentum" of the scalar field 
$P_{\phi }=(\partial L/\partial \dot{\phi })=\dot{\phi }$. 
Substituting into 
Eq. (5) and solving for $H$, one can bring Eq. (6) into the form 
\bea 
(1+\frac{1}{2}c_{0}^{2})\dot{F}&+&\frac{3\kp }{2}
\left[1+\frac{1}{2}c_{0}^{2}\right]^{1/2}c_{0}F^{2}
-\frac{\dot{c_{0}}}{c_{0}}F=0\nn \\
F\equiv \dot{\phi }, &\; &\; \; \; \; 
\kp ^{2}\equiv \frac{8\pi G}{3}.
\eea 
Setting 
\be 
F=[1+(f/2)c_{0}^{2}]^{-1},
\ee 
with $f$ a positive constant, 
one obtains the following equation to be satisfied 
by $c_{0}(t)$, 
\be 
\dot{c_{0}}=
\frac{3\kp }{2}\frac{c_{0}^{2}[1+(1/2)c_{0}^{2}]^{1/2}}
{[1+(3/2)fc_{0}^{2}+(f/2)c_{0}^{4}]}.
\ee 
The ansatz of Eq. (15) essentially amounts to a choice of 
an equation of state, because this determines the function 
$c_{0}(t)$ and so an equation of state, through the time dependent 
version of Eq. (11).\\
The initial condition is $c_{0}(t=0)=c_{0I}>0$. Now 
inflation will last as long as $c_{0}^{2}(t)<1$. The function 
$c_{0}(t)$ is {\it continuous, strictly increasing} 
(i.e. $\dot{c_{0}}>0$), therefore is invertible in the range 
of $t\in [0,+\infty )$ with $c_{0I}\leq c_{0}(t)\leq c_{0F}$. 
The behaviour of $c_{0}(t)$ will be investigated numerically 
in the next section. The equation of state is given again 
by the time-dependent version of Eq. (11). This interpolates 
between the value -1 for the quantum vacuum (in the early stages) 
and the asymptotic value +1 (when $c_{0}(t)\gg 1$) for the stiff 
matter.\\
The relevant interesting asymptotic limits are 
\bea 
(i)\; \; \; \; \dot{c_{0}}=\frac{3\kp }{2}c_{0}^{2}
\; \; \; (c_{0}\ll 1)\nn \\
(ii)\; \; \; \; \dot{c_{0}}=\frac{3\kp }{f\sqrt{2}c_{0}}\; \; \; (c_{0}\gg 1).
\eea 
In the first case we obtain 
($0\leq t<t_{0}\equiv (3\kp c_{0I}/2)^{-1}$)
\bea 
c_{0}(t)&=&\frac{c_{0I}}{1-3c_{0I}\kp t/2}\nn \\
\phi (t)=t&-&\frac{\sqrt{2}}{3\kp }
Arctan\left[\frac{\sqrt{2}}{c_{0I}}(\frac{3}{2}\kp c_{0I}t-1)\right]
-\frac{\sqrt{2}}{3\kp }
Arctan\left[\frac{\sqrt{2}}{c_{0I}}\right]
\nn \\
H&=&\kp 
\left[\frac{\dot{\phi }^{2}}{2}+
\frac{\dot{\phi }^{2}}{c_{0}^{2}}\right]^{1/2}.
\eea 
This yields the scale factor which is regular at $t\simeq 0$ and has the 
de Sitter form 
\be 
a(t)\simeq a_{0}exp\left(\kp \sqrt{2}
\frac{t}{c_{0I}\sqrt{2+c_{0I}^{2}}}\right).
\ee 
The number of e-folds of inflation, with $c_{0I}=0.012$, 
is as large as $c_{0I}^{-1}$.\\
In the second case 
we obtain 
\bea 
c_{0}^{2}(t)&=&c_{0I}^{2}+\frac{3\kp }{f\sqrt{2}}t\nn \\
H&=&\frac{\kp }{c_{0}(t)}
\frac{\left(1+\frac{1}{2}c_{0}^{2}\right)^{1/2}}
{\left(1+\frac{f}{2}c_{0}^{2}\right)}.
\eea 
This yields a Hubble parameter which depends on time 
as $H(2/3)t^{-1}$ and one 
obtains, for all the members of this class 
(for every positive value of $f$) 
the matter dominated Universe at late evolution times\\
\be 
a(t)=a_{0}t^{2/3}. 
\ee 
We summarize for the sake of completeness the method presented here. 
One integrates Eq. (16) for $c_{0}(t)$, with initial condition 
$c_{0I}$. Then through Eq. (15) one obtains $\phi =\phi (t)$ which is 
invertible (through $F=\dot{\phi }>0, \forall t\geq 0$) 
for $t=t(\phi )$. Therefore from Eq. (8) one obtains, 
in principle, the functional form of the potential $V=V(\phi )$. 
Finally with the aid of Eq. (2) one obtains the scalar density $\rho $ 
and from Eq. (5) one obtains the scale factor $a=a(t)$. Thus this 
is a straightforward procedure to obtain an explicit, time dependent, 
solution to the field equations Eq. (2) and Eq. (5-6), along with 
the equation of state, Eq. (11). 

The main advantage of these models is that the scale factors 
are nonsingular at the time $t=0$. The scalar curvature is 
finite and therefore these are suitable for the Pre-Big-Bang 
(PBB) scenarios [5,7]. 
Also the transition from 
the inflationary to the deccelerating era occurs smoothly, 
when $c_{0}(t)>1$, 
at a time controlled by the initial value $c_{0I}$. Also in 
a natural manner, in one of these models the matter dominated 
Universe occurs in the late stages of the evolution. 
These aspects are investigated numerically in the next section. 

\section*{IV. Numerical Results}\
In this section we present the results of the numerical treatment 
of the set of differential equations as they appear in Eqs. (15), (16) 
and Eq. (18c). The initial conditions are $0<c_{0}(t=0)=c_{0I}\ll 1$, 
while for the scalar field $\phi (t=0)=0$. That is we use Eq. (16) and 
\bea 
\dot{\phi }&=&\frac{1}{[1+(f/2)c_{0}^{2}(t)]} \\
\dot{a}&=&a\frac{\kp \dot{\phi }}{c_{0}}\sqrt{(1+(1/2)c_{0}^{2}(t)}.
\eea 
For the scale factor we use $a(t=0)=a_{0}>0$. We use units where 
$\hbar =c=1$ so that $t_{Planck}=(\hbar G/c^{5})\sim \kp $. 
The results for the scale
factors are plotted in Figs. (1)-(2).  In these plots the scale factors
are normalized to their initial values which are finite, therefore the
plots cross the vertical axis at 1, for $t=0$. Finally through the
numerical integration of the system, for the range of initial conditions 
$0.01\leq c_{0I}\leq 0.05$, the dependence of the time that inflation 
ends, $t_{*}$ (in Planck units), on $c_{0I}$ has the form  
$t_{*}=0.063c_{0I}^{-1/2}$.
\newpage 

\begin {figure}[h]
\begin{center}
\setlength{\unitlength}{0.240900pt}
\ifx\plotpoint\undefined\newsavebox{\plotpoint}\fi
\sbox{\plotpoint}{\rule[-0.200pt]{0.400pt}{0.400pt}}%
\begin{picture}(1500,900)(0,0)
\font\gnuplot=cmr10 at 10pt
\gnuplot
\sbox{\plotpoint}{\rule[-0.200pt]{0.400pt}{0.400pt}}%
\put(282.0,123.0){\rule[-0.200pt]{4.818pt}{0.400pt}}
\put(262,123){\makebox(0,0)[r]{0}}
\put(1419.0,123.0){\rule[-0.200pt]{4.818pt}{0.400pt}}
\put(282.0,232.0){\rule[-0.200pt]{4.818pt}{0.400pt}}
\put(262,232){\makebox(0,0)[r]{5e+37}}
\put(1419.0,232.0){\rule[-0.200pt]{4.818pt}{0.400pt}}
\put(282.0,341.0){\rule[-0.200pt]{4.818pt}{0.400pt}}
\put(262,341){\makebox(0,0)[r]{1e+38}}
\put(1419.0,341.0){\rule[-0.200pt]{4.818pt}{0.400pt}}
\put(282.0,450.0){\rule[-0.200pt]{4.818pt}{0.400pt}}
\put(262,450){\makebox(0,0)[r]{1.5e+38}}
\put(1419.0,450.0){\rule[-0.200pt]{4.818pt}{0.400pt}}
\put(282.0,559.0){\rule[-0.200pt]{4.818pt}{0.400pt}}
\put(262,559){\makebox(0,0)[r]{2e+38}}
\put(1419.0,559.0){\rule[-0.200pt]{4.818pt}{0.400pt}}
\put(282.0,668.0){\rule[-0.200pt]{4.818pt}{0.400pt}}
\put(262,668){\makebox(0,0)[r]{2.5e+38}}
\put(1419.0,668.0){\rule[-0.200pt]{4.818pt}{0.400pt}}
\put(282.0,777.0){\rule[-0.200pt]{4.818pt}{0.400pt}}
\put(262,777){\makebox(0,0)[r]{3e+38}}
\put(1419.0,777.0){\rule[-0.200pt]{4.818pt}{0.400pt}}
\put(282.0,123.0){\rule[-0.200pt]{0.400pt}{4.818pt}}
\put(282,82){\makebox(0,0){0}}
\put(282.0,757.0){\rule[-0.200pt]{0.400pt}{4.818pt}}
\put(398.0,123.0){\rule[-0.200pt]{0.400pt}{4.818pt}}
\put(398,82){\makebox(0,0){10}}
\put(398.0,757.0){\rule[-0.200pt]{0.400pt}{4.818pt}}
\put(513.0,123.0){\rule[-0.200pt]{0.400pt}{4.818pt}}
\put(513,82){\makebox(0,0){20}}
\put(513.0,757.0){\rule[-0.200pt]{0.400pt}{4.818pt}}
\put(629.0,123.0){\rule[-0.200pt]{0.400pt}{4.818pt}}
\put(629,82){\makebox(0,0){30}}
\put(629.0,757.0){\rule[-0.200pt]{0.400pt}{4.818pt}}
\put(745.0,123.0){\rule[-0.200pt]{0.400pt}{4.818pt}}
\put(745,82){\makebox(0,0){40}}
\put(745.0,757.0){\rule[-0.200pt]{0.400pt}{4.818pt}}
\put(861.0,123.0){\rule[-0.200pt]{0.400pt}{4.818pt}}
\put(861,82){\makebox(0,0){50}}
\put(861.0,757.0){\rule[-0.200pt]{0.400pt}{4.818pt}}
\put(976.0,123.0){\rule[-0.200pt]{0.400pt}{4.818pt}}
\put(976,82){\makebox(0,0){60}}
\put(976.0,757.0){\rule[-0.200pt]{0.400pt}{4.818pt}}
\put(1092.0,123.0){\rule[-0.200pt]{0.400pt}{4.818pt}}
\put(1092,82){\makebox(0,0){70}}
\put(1092.0,757.0){\rule[-0.200pt]{0.400pt}{4.818pt}}
\put(1208.0,123.0){\rule[-0.200pt]{0.400pt}{4.818pt}}
\put(1208,82){\makebox(0,0){80}}
\put(1208.0,757.0){\rule[-0.200pt]{0.400pt}{4.818pt}}
\put(1323.0,123.0){\rule[-0.200pt]{0.400pt}{4.818pt}}
\put(1323,82){\makebox(0,0){90}}
\put(1323.0,757.0){\rule[-0.200pt]{0.400pt}{4.818pt}}
\put(1439.0,123.0){\rule[-0.200pt]{0.400pt}{4.818pt}}
\put(1439,82){\makebox(0,0){100}}
\put(1439.0,757.0){\rule[-0.200pt]{0.400pt}{4.818pt}}
\put(282.0,123.0){\rule[-0.200pt]{278.721pt}{0.400pt}}
\put(1439.0,123.0){\rule[-0.200pt]{0.400pt}{157.549pt}}
\put(282.0,777.0){\rule[-0.200pt]{278.721pt}{0.400pt}}
\put(40,450){\makebox(0,0){(a(t)/$a_{0}$)}}
\put(860,21){\makebox(0,0){($t/t_{Planck}$)}}
\put(860,839){\makebox(0,0){Initial Condition $c_{0I}=0.012$}}
\put(340,269){\makebox(0,0)[l]{------- End of Inflation}}
\put(282.0,123.0){\rule[-0.200pt]{0.400pt}{157.549pt}}
\put(282,123){\usebox{\plotpoint}}
\put(282,123){\usebox{\plotpoint}}
\multiput(317.58,123.00)(0.492,1.345){19}{\rule{0.118pt}{1.155pt}}
\multiput(316.17,123.00)(11.000,26.604){2}{\rule{0.400pt}{0.577pt}}
\multiput(328.58,152.00)(0.492,5.106){21}{\rule{0.119pt}{4.067pt}}
\multiput(327.17,152.00)(12.000,110.559){2}{\rule{0.400pt}{2.033pt}}
\multiput(340.58,271.00)(0.492,4.081){19}{\rule{0.118pt}{3.264pt}}
\multiput(339.17,271.00)(11.000,80.226){2}{\rule{0.400pt}{1.632pt}}
\multiput(351.58,358.00)(0.492,2.349){21}{\rule{0.119pt}{1.933pt}}
\multiput(350.17,358.00)(12.000,50.987){2}{\rule{0.400pt}{0.967pt}}
\multiput(363.58,413.00)(0.492,1.659){21}{\rule{0.119pt}{1.400pt}}
\multiput(362.17,413.00)(12.000,36.094){2}{\rule{0.400pt}{0.700pt}}
\multiput(375.58,452.00)(0.492,1.345){19}{\rule{0.118pt}{1.155pt}}
\multiput(374.17,452.00)(11.000,26.604){2}{\rule{0.400pt}{0.577pt}}
\multiput(386.58,481.00)(0.492,0.927){21}{\rule{0.119pt}{0.833pt}}
\multiput(385.17,481.00)(12.000,20.270){2}{\rule{0.400pt}{0.417pt}}
\multiput(398.58,503.00)(0.492,0.826){19}{\rule{0.118pt}{0.755pt}}
\multiput(397.17,503.00)(11.000,16.434){2}{\rule{0.400pt}{0.377pt}}
\multiput(409.58,521.00)(0.492,0.625){21}{\rule{0.119pt}{0.600pt}}
\multiput(408.17,521.00)(12.000,13.755){2}{\rule{0.400pt}{0.300pt}}
\multiput(421.58,536.00)(0.492,0.590){19}{\rule{0.118pt}{0.573pt}}
\multiput(420.17,536.00)(11.000,11.811){2}{\rule{0.400pt}{0.286pt}}
\multiput(432.00,549.58)(0.543,0.492){19}{\rule{0.536pt}{0.118pt}}
\multiput(432.00,548.17)(10.887,11.000){2}{\rule{0.268pt}{0.400pt}}
\multiput(444.00,560.58)(0.600,0.491){17}{\rule{0.580pt}{0.118pt}}
\multiput(444.00,559.17)(10.796,10.000){2}{\rule{0.290pt}{0.400pt}}
\multiput(456.00,570.59)(0.692,0.488){13}{\rule{0.650pt}{0.117pt}}
\multiput(456.00,569.17)(9.651,8.000){2}{\rule{0.325pt}{0.400pt}}
\multiput(467.00,578.59)(0.758,0.488){13}{\rule{0.700pt}{0.117pt}}
\multiput(467.00,577.17)(10.547,8.000){2}{\rule{0.350pt}{0.400pt}}
\multiput(479.00,586.59)(0.798,0.485){11}{\rule{0.729pt}{0.117pt}}
\multiput(479.00,585.17)(9.488,7.000){2}{\rule{0.364pt}{0.400pt}}
\multiput(490.00,593.59)(1.033,0.482){9}{\rule{0.900pt}{0.116pt}}
\multiput(490.00,592.17)(10.132,6.000){2}{\rule{0.450pt}{0.400pt}}
\multiput(502.00,599.59)(0.943,0.482){9}{\rule{0.833pt}{0.116pt}}
\multiput(502.00,598.17)(9.270,6.000){2}{\rule{0.417pt}{0.400pt}}
\multiput(513.00,605.59)(1.267,0.477){7}{\rule{1.060pt}{0.115pt}}
\multiput(513.00,604.17)(9.800,5.000){2}{\rule{0.530pt}{0.400pt}}
\multiput(525.00,610.59)(1.267,0.477){7}{\rule{1.060pt}{0.115pt}}
\multiput(525.00,609.17)(9.800,5.000){2}{\rule{0.530pt}{0.400pt}}
\multiput(537.00,615.59)(1.155,0.477){7}{\rule{0.980pt}{0.115pt}}
\multiput(537.00,614.17)(8.966,5.000){2}{\rule{0.490pt}{0.400pt}}
\multiput(548.00,620.60)(1.651,0.468){5}{\rule{1.300pt}{0.113pt}}
\multiput(548.00,619.17)(9.302,4.000){2}{\rule{0.650pt}{0.400pt}}
\multiput(560.00,624.60)(1.505,0.468){5}{\rule{1.200pt}{0.113pt}}
\multiput(560.00,623.17)(8.509,4.000){2}{\rule{0.600pt}{0.400pt}}
\multiput(571.00,628.61)(2.472,0.447){3}{\rule{1.700pt}{0.108pt}}
\multiput(571.00,627.17)(8.472,3.000){2}{\rule{0.850pt}{0.400pt}}
\multiput(583.00,631.60)(1.505,0.468){5}{\rule{1.200pt}{0.113pt}}
\multiput(583.00,630.17)(8.509,4.000){2}{\rule{0.600pt}{0.400pt}}
\multiput(594.00,635.61)(2.472,0.447){3}{\rule{1.700pt}{0.108pt}}
\multiput(594.00,634.17)(8.472,3.000){2}{\rule{0.850pt}{0.400pt}}
\multiput(606.00,638.61)(2.472,0.447){3}{\rule{1.700pt}{0.108pt}}
\multiput(606.00,637.17)(8.472,3.000){2}{\rule{0.850pt}{0.400pt}}
\multiput(618.00,641.61)(2.248,0.447){3}{\rule{1.567pt}{0.108pt}}
\multiput(618.00,640.17)(7.748,3.000){2}{\rule{0.783pt}{0.400pt}}
\multiput(629.00,644.61)(2.472,0.447){3}{\rule{1.700pt}{0.108pt}}
\multiput(629.00,643.17)(8.472,3.000){2}{\rule{0.850pt}{0.400pt}}
\put(641,647.17){\rule{2.300pt}{0.400pt}}
\multiput(641.00,646.17)(6.226,2.000){2}{\rule{1.150pt}{0.400pt}}
\multiput(652.00,649.61)(2.472,0.447){3}{\rule{1.700pt}{0.108pt}}
\multiput(652.00,648.17)(8.472,3.000){2}{\rule{0.850pt}{0.400pt}}
\put(664,652.17){\rule{2.300pt}{0.400pt}}
\multiput(664.00,651.17)(6.226,2.000){2}{\rule{1.150pt}{0.400pt}}
\multiput(675.00,654.61)(2.472,0.447){3}{\rule{1.700pt}{0.108pt}}
\multiput(675.00,653.17)(8.472,3.000){2}{\rule{0.850pt}{0.400pt}}
\put(687,657.17){\rule{2.500pt}{0.400pt}}
\multiput(687.00,656.17)(6.811,2.000){2}{\rule{1.250pt}{0.400pt}}
\put(699,659.17){\rule{2.300pt}{0.400pt}}
\multiput(699.00,658.17)(6.226,2.000){2}{\rule{1.150pt}{0.400pt}}
\put(710,661.17){\rule{2.500pt}{0.400pt}}
\multiput(710.00,660.17)(6.811,2.000){2}{\rule{1.250pt}{0.400pt}}
\put(722,663.17){\rule{2.300pt}{0.400pt}}
\multiput(722.00,662.17)(6.226,2.000){2}{\rule{1.150pt}{0.400pt}}
\put(733,665.17){\rule{2.500pt}{0.400pt}}
\multiput(733.00,664.17)(6.811,2.000){2}{\rule{1.250pt}{0.400pt}}
\put(745,667.17){\rule{2.300pt}{0.400pt}}
\multiput(745.00,666.17)(6.226,2.000){2}{\rule{1.150pt}{0.400pt}}
\put(756,668.67){\rule{2.891pt}{0.400pt}}
\multiput(756.00,668.17)(6.000,1.000){2}{\rule{1.445pt}{0.400pt}}
\put(768,670.17){\rule{2.500pt}{0.400pt}}
\multiput(768.00,669.17)(6.811,2.000){2}{\rule{1.250pt}{0.400pt}}
\put(780,672.17){\rule{2.300pt}{0.400pt}}
\multiput(780.00,671.17)(6.226,2.000){2}{\rule{1.150pt}{0.400pt}}
\put(791,673.67){\rule{2.891pt}{0.400pt}}
\multiput(791.00,673.17)(6.000,1.000){2}{\rule{1.445pt}{0.400pt}}
\put(803,675.17){\rule{2.300pt}{0.400pt}}
\multiput(803.00,674.17)(6.226,2.000){2}{\rule{1.150pt}{0.400pt}}
\put(814,676.67){\rule{2.891pt}{0.400pt}}
\multiput(814.00,676.17)(6.000,1.000){2}{\rule{1.445pt}{0.400pt}}
\put(826,678.17){\rule{2.300pt}{0.400pt}}
\multiput(826.00,677.17)(6.226,2.000){2}{\rule{1.150pt}{0.400pt}}
\put(837,679.67){\rule{2.891pt}{0.400pt}}
\multiput(837.00,679.17)(6.000,1.000){2}{\rule{1.445pt}{0.400pt}}
\put(849,680.67){\rule{2.891pt}{0.400pt}}
\multiput(849.00,680.17)(6.000,1.000){2}{\rule{1.445pt}{0.400pt}}
\put(861,682.17){\rule{2.300pt}{0.400pt}}
\multiput(861.00,681.17)(6.226,2.000){2}{\rule{1.150pt}{0.400pt}}
\put(872,683.67){\rule{2.891pt}{0.400pt}}
\multiput(872.00,683.17)(6.000,1.000){2}{\rule{1.445pt}{0.400pt}}
\put(884,684.67){\rule{2.650pt}{0.400pt}}
\multiput(884.00,684.17)(5.500,1.000){2}{\rule{1.325pt}{0.400pt}}
\put(895,686.17){\rule{2.500pt}{0.400pt}}
\multiput(895.00,685.17)(6.811,2.000){2}{\rule{1.250pt}{0.400pt}}
\put(907,687.67){\rule{2.650pt}{0.400pt}}
\multiput(907.00,687.17)(5.500,1.000){2}{\rule{1.325pt}{0.400pt}}
\put(918,688.67){\rule{2.891pt}{0.400pt}}
\multiput(918.00,688.17)(6.000,1.000){2}{\rule{1.445pt}{0.400pt}}
\put(930,689.67){\rule{2.650pt}{0.400pt}}
\multiput(930.00,689.17)(5.500,1.000){2}{\rule{1.325pt}{0.400pt}}
\put(941,690.67){\rule{2.891pt}{0.400pt}}
\multiput(941.00,690.17)(6.000,1.000){2}{\rule{1.445pt}{0.400pt}}
\put(953,691.67){\rule{2.891pt}{0.400pt}}
\multiput(953.00,691.17)(6.000,1.000){2}{\rule{1.445pt}{0.400pt}}
\put(965,692.67){\rule{2.650pt}{0.400pt}}
\multiput(965.00,692.17)(5.500,1.000){2}{\rule{1.325pt}{0.400pt}}
\put(976,693.67){\rule{2.891pt}{0.400pt}}
\multiput(976.00,693.17)(6.000,1.000){2}{\rule{1.445pt}{0.400pt}}
\put(988,694.67){\rule{2.650pt}{0.400pt}}
\multiput(988.00,694.17)(5.500,1.000){2}{\rule{1.325pt}{0.400pt}}
\put(999,695.67){\rule{2.891pt}{0.400pt}}
\multiput(999.00,695.17)(6.000,1.000){2}{\rule{1.445pt}{0.400pt}}
\put(1011,696.67){\rule{2.650pt}{0.400pt}}
\multiput(1011.00,696.17)(5.500,1.000){2}{\rule{1.325pt}{0.400pt}}
\put(1022,697.67){\rule{2.891pt}{0.400pt}}
\multiput(1022.00,697.17)(6.000,1.000){2}{\rule{1.445pt}{0.400pt}}
\put(1034,698.67){\rule{2.891pt}{0.400pt}}
\multiput(1034.00,698.17)(6.000,1.000){2}{\rule{1.445pt}{0.400pt}}
\put(1046,699.67){\rule{2.650pt}{0.400pt}}
\multiput(1046.00,699.17)(5.500,1.000){2}{\rule{1.325pt}{0.400pt}}
\put(1057,700.67){\rule{2.891pt}{0.400pt}}
\multiput(1057.00,700.17)(6.000,1.000){2}{\rule{1.445pt}{0.400pt}}
\put(282.0,123.0){\rule[-0.200pt]{8.431pt}{0.400pt}}
\put(1080,701.67){\rule{2.891pt}{0.400pt}}
\multiput(1080.00,701.17)(6.000,1.000){2}{\rule{1.445pt}{0.400pt}}
\put(1092,702.67){\rule{2.650pt}{0.400pt}}
\multiput(1092.00,702.17)(5.500,1.000){2}{\rule{1.325pt}{0.400pt}}
\put(1103,703.67){\rule{2.891pt}{0.400pt}}
\multiput(1103.00,703.17)(6.000,1.000){2}{\rule{1.445pt}{0.400pt}}
\put(1115,704.67){\rule{2.891pt}{0.400pt}}
\multiput(1115.00,704.17)(6.000,1.000){2}{\rule{1.445pt}{0.400pt}}
\put(1069.0,702.0){\rule[-0.200pt]{2.650pt}{0.400pt}}
\put(1138,705.67){\rule{2.891pt}{0.400pt}}
\multiput(1138.00,705.17)(6.000,1.000){2}{\rule{1.445pt}{0.400pt}}
\put(1150,706.67){\rule{2.650pt}{0.400pt}}
\multiput(1150.00,706.17)(5.500,1.000){2}{\rule{1.325pt}{0.400pt}}
\put(1161,707.67){\rule{2.891pt}{0.400pt}}
\multiput(1161.00,707.17)(6.000,1.000){2}{\rule{1.445pt}{0.400pt}}
\put(1127.0,706.0){\rule[-0.200pt]{2.650pt}{0.400pt}}
\put(1184,708.67){\rule{2.891pt}{0.400pt}}
\multiput(1184.00,708.17)(6.000,1.000){2}{\rule{1.445pt}{0.400pt}}
\put(1196,709.67){\rule{2.891pt}{0.400pt}}
\multiput(1196.00,709.17)(6.000,1.000){2}{\rule{1.445pt}{0.400pt}}
\put(1173.0,709.0){\rule[-0.200pt]{2.650pt}{0.400pt}}
\put(1219,710.67){\rule{2.891pt}{0.400pt}}
\multiput(1219.00,710.17)(6.000,1.000){2}{\rule{1.445pt}{0.400pt}}
\put(1231,711.67){\rule{2.650pt}{0.400pt}}
\multiput(1231.00,711.17)(5.500,1.000){2}{\rule{1.325pt}{0.400pt}}
\put(1208.0,711.0){\rule[-0.200pt]{2.650pt}{0.400pt}}
\put(1254,712.67){\rule{2.650pt}{0.400pt}}
\multiput(1254.00,712.17)(5.500,1.000){2}{\rule{1.325pt}{0.400pt}}
\put(1265,713.67){\rule{2.891pt}{0.400pt}}
\multiput(1265.00,713.17)(6.000,1.000){2}{\rule{1.445pt}{0.400pt}}
\put(1242.0,713.0){\rule[-0.200pt]{2.891pt}{0.400pt}}
\put(1289,714.67){\rule{2.650pt}{0.400pt}}
\multiput(1289.00,714.17)(5.500,1.000){2}{\rule{1.325pt}{0.400pt}}
\put(1277.0,715.0){\rule[-0.200pt]{2.891pt}{0.400pt}}
\put(1312,715.67){\rule{2.650pt}{0.400pt}}
\multiput(1312.00,715.17)(5.500,1.000){2}{\rule{1.325pt}{0.400pt}}
\put(1323,716.67){\rule{2.891pt}{0.400pt}}
\multiput(1323.00,716.17)(6.000,1.000){2}{\rule{1.445pt}{0.400pt}}
\put(1300.0,716.0){\rule[-0.200pt]{2.891pt}{0.400pt}}
\put(1346,717.67){\rule{2.891pt}{0.400pt}}
\multiput(1346.00,717.17)(6.000,1.000){2}{\rule{1.445pt}{0.400pt}}
\put(1335.0,718.0){\rule[-0.200pt]{2.650pt}{0.400pt}}
\put(1370,718.67){\rule{2.650pt}{0.400pt}}
\multiput(1370.00,718.17)(5.500,1.000){2}{\rule{1.325pt}{0.400pt}}
\put(1358.0,719.0){\rule[-0.200pt]{2.891pt}{0.400pt}}
\put(1393,719.67){\rule{2.650pt}{0.400pt}}
\multiput(1393.00,719.17)(5.500,1.000){2}{\rule{1.325pt}{0.400pt}}
\put(1381.0,720.0){\rule[-0.200pt]{2.891pt}{0.400pt}}
\put(1416,720.67){\rule{2.650pt}{0.400pt}}
\multiput(1416.00,720.17)(5.500,1.000){2}{\rule{1.325pt}{0.400pt}}
\put(1404.0,721.0){\rule[-0.200pt]{2.891pt}{0.400pt}}
\end{picture}

\end{center}
\caption{Normalized Scale factor for $c_{0I}=0.012$}

\begin{center}
\input{g0015.tex}
\end{center}
\caption{Normalized Scale factor for $c_{0I}=0.015$}
\end{figure} 

\section*{VI. Conclusions and Discussion}\
In this paper we have presented some new solutions 
to the scalar field-driven inflation. The main interesting 
features of these models can be summarized as follows:  
These are 
exact solutions and no use of the slow-roll approximation 
has been made. All the terms in the scalar field equation of 
motion (Klein-Gordon equation) are present. Also in this 
class of models the scale factor is nonsingular for all 
positive values of the evolution time. Moreover the transition 
from the inflationary to  
the deccelerating phase is smooth and occurs in a 
cosmic time that depends on the initial value of the 
canonical momentum of the scalar field. Finally for all the members 
of this class there exists the asymptotic transition to the 
matter dominated Universe. \\
The number of e-folds of 
inflation is high enough to solve the well known problems of the 
standard model, by a suitable choice of the initial value of 
the scalar field, although improvement is certainly possible. This 
is currently under investigation. \\

\section*{Acknowledgements}
The authors would like to thank Dr. Christos Vozikis 
for valuable assistance in the computational parts of 
this paper.

\end{document}